\documentclass[onecolumn,showpacs,prb,aps]{revtex4}
\usepackage{graphicx}

\begin{document}

\title{
Impurity effects on optical response in a finite band electronic
system coupled to phonons\\
}

\author{Anton~Knigavko}
\email{anton.knigavko@brocku.ca}
\affiliation{Department of Physics, Brock University,
St.~Catharines, Ontario, Canada, L2S 3A1}

\author{J.~P.~Carbotte}
\email{carbotte@mcmaster.ca}
\affiliation{Department of Physics and Astronomy, McMaster
University, Hamilton, Ontario, Canada, L8S 4M1}

\date{\today}

\begin{abstract}
The concepts, which have traditionally been useful in understanding
the effects of the electron--phonon interaction in optical
spectroscopy, are based on insights obtained within the infinite
electronic band approximation and no longer apply in finite band
metals. Impurity and phonon contributions to electron scattering are
not additive and the apparent strength of the coupling to the phonon
degrees of freedom is substantially reduced with increased elastic
scattering. The optical mass renormalization changes sign with
increasing frequency and the optical scattering rate never reaches
its high frequency quasiparticle value which itself is also reduced
below its infinite band value.
\end{abstract}
\maketitle

\section{Introduction}

Many of the physical insights that have guided the interpretation of
data on effects of the electron--phonon interaction in metals are
based on an infinite band model with a constant featureless
electronic density of states (EDOS) \cite{mars-book}. In the
eighties there appeared several studies, mainly motivated by the
physics of the A15 compounds, which took account of energy
dependence in the EDOS around the chemical potential
\cite{mitrovic83,pickett82,mitrovic-thesis81}.
Band structure calculations for the A15
showed peaks in the EDOS with variations on the energy scale of
$50$ meV. A variety of experiments also showed sensitivity of
properties to disorder. For example, disordered
Mo$_3$Ge has a higher value of superconducting critical
temperature than its crystalline counterpart. This is naturally
explained if the chemical potential in ordered metallic Mo$_3$Ge
falls in a valley of the EDOS. Radiation damage then fills this
valley and leads to an increase in EDOS at the Fermi energy and a
higher value of T$_c$.

More recently several authors have considered a different but
related effect, namely a finite band
\cite{cappelluti03,dogan03,knigavko05,knigavko05-2}.
One experimental realization of this situation is the fulleride
compounds M$_3$C$_{60}$ ($M$ -- an alkali metal),
where band structure calculations \cite{gelfand94}
show narrow band with the width $W$ of the order of 1 eV,
while the phonon spectrum extends up to about 200 meV
in some cases. Physical consequences
brought about by the finite band can be studied in the frame of a
simplified (particle--hole symmetric) model with  a constant $N_0$
with a cut off applied at $\pm W/2$ where $W$ the band width related
to $N_0$ by $N_0=1/W$.
 A somewhat surprising result of such studies
is that, even for rather wide bands ($W$ of order a few eV) certain
aspects associated with the effect of the electron--phonon
interaction are profoundly modified as compared to the corresponding
infinite band behavior. For example, in an infinite band with
constant electronic density of states, the electron--phonon
interaction leaves $N_0(\omega)$ unaltered and no phonon structure
appears in the dressed normal state EDOS. To see phonon structure it
is necessary to go to the superconducting state which develops a gap
and consequently a non constant EDOS. However if a cut off is
applied to the constant $N_0(\omega)$, then phonon structure appears
in the dressed quasiparticle density of states as it does in the
superconducting state and also in any case when EDOS is non constant
around the Fermi energy. The phonon structure which appears in the
dressed EDOS is surprisingly significant in magnitude even for
modest value of the electron--phonon mass renormalization parameter
$\lambda$. Mathematically the self energy must be solved for self
consistently when a finite band cut off is introduced. This
contrasts with the infinite band case where the bare Green's
function can be employed in the self energy expression
\cite{englesberg63}.
Self consistency leads to a smearing of the band edge region as well
as a widening of the band. As the total number of states in the
electronic density of state must remain constant, this transfer of
spectral weight to higher energies beyond the bare EDOS cut off,
implies that it must correspondingly be reduced at smaller energies
and the details of this reduction depend significantly on the phonon
energy scale and coupling strength to the various phonon modes.
The effect of widening of a finite electronic band due to the
electron--phonon interaction  has been observed and discussed
previously by Liechtenstein {\it et al} \cite{liechtenstein96}
in the context of fulleride compounds.

There are many qualitative  changes in electron--phonon
renormalization effects which have their origin in finite bands. For
example the real part of the electronic self energy
$\Sigma_1(\omega)$ for $\omega \geq 0$ is everywhere negative in an
infinite band and decays to zero beyond a few times the maximum
phonon energy, which we denote $\omega_D$. Thus the electronic effective mass is
always increased by the electron--phonon interaction and returns to
its bare mass value from above at a few times $\omega_D$. By
contrast for a finite band, as described in
Ref.~\onlinecite{cappelluti03}, $\Sigma_1(\omega)$ change sign as
$\omega$ increases and the renormalized mass at high $\omega$ can
actually be smaller than the bare band mass. This is an example of
qualitative change brought about in the electronic self energy by
finite band effects. Others are described in the recent paper of
Cappelluti and Pietronero \cite{cappelluti03}, who also considered
the effect of impurities.

In this paper we consider optical properties with particular
emphasis on the combined effect of temperature and impurity
scattering in a finite band electron--phonon system. In Section II
we provide a brief summary of the formalism needed to compute the
electron self energy $\Sigma$ vs $\omega$ for a system of electrons
coupled both to phonons and to impurities. We also present analytic
formulas which apply in the non selfconsistent
approximation. They will prove useful for interpretation of the
numerical results. The optical conductivity
without vertex corrections follows from the Kubo formula for the
current--current correlation function. The optical self energy, or
the memory function, is introduced and related to the complex
optical conductivity $\sigma(\omega)$.
We summarize some known approximate but analytic formulas for the
optical scattering rate and effective mass renormalization which
have been found useful in past studies  related to infinite
(very wide) electronic bands. These formulas
appropriately modified in the context of finite bands are applied
to obtain  a description of the non selfconsistent approximation,
which provide insight into the various features found in numerical
solution of the full equations. Two models for the
electron--phonon spectral density $\alpha^2 F(\omega)$ are
introduced. For definiteness both are based on the specific phonon
spectrum of $K_3C_{60}$. One consists of three delta functions
suitably chosen to mimic  the real spectrum while the other one uses
truncated Lorentzians instead of delta functions  to help
understand the modification brought about when the extended nature of
real spectra is accounted for. In Section III we describe results
for the case of a rather wide band and the three delta function
model for $\alpha^2 F(\omega)$ with a modest value of
$\lambda=0.71$. We start with a discussion of the dressed electronic
density of states with emphasis on temperature and impurity effects.
Then the memory function is analyzed and compared with the
quasiparticle self energy, the differences arising from finite
band effects are emphasized. In Section IV  we  present the results
for an extended electron--phonon spectrum, increasing the spectral
$\lambda$ and decreasing the width of the band. Section V is
our conclusions.

\section{Formalism}

\subsection{The electronic self energy and the renormalized EDOS}

The central quantity of our problem is the electronic self energy
$\Sigma (z)=\Sigma _{1}(z)+{\rm i}\Sigma _{2}(z)$.
It is calculated from the Migdal equations formulated in the
mixed real--imaginary axis representation \cite{cappelluti03,marsiglio88}:
\begin{eqnarray}
\Sigma (z) &=&\Gamma \, \eta (z)+T\sum_{m=-\infty }^{+\infty }
\lambda(z-{\rm i}\omega _{m})\eta ({\rm i}\omega _{m})
\nonumber \\
&+& \int_{0}^{\infty }{\rm d}\omega \,\alpha ^{2}F(\omega )\left\{ \left[
f(\omega -z)+n(\omega )\right] \eta (z-\omega )
+ \left[ f(\omega +z)+n(\omega )\right] \eta (z+\omega )\right\} ,
\label{self-energy} \\
\lambda (z) &=&\int_{0}^{\infty }{\rm d}\omega \,\alpha ^{2}F(\omega )
\frac{2\omega }{\omega^{2}-z^{2}} ,
\label{lambda} \\
\eta (z) &=&\int_{-\infty }^{\infty }{\rm d}\xi \,\frac{N_{0}(\xi )}{N_{0}(0)}%
\frac{1}{z-\xi -\Sigma (z)} ,
\label{eta}
\end{eqnarray}
where $\omega _{m}=\pi T(2m-1),$ $m\in Z$ are the fermionic Matsubara
frequencies, and $f(\omega )$ and $n(\omega )$ are the Fermi and Bose distribution
functions respectively. The electron--phonon interaction is specified in
terms of the electron--phonon spectral function $\alpha ^{2}F(\omega )$ (the
Eliashberg function). The parameter $\Gamma $, which has the meaning of an impurity
scattering rate, specifies the strength of the interaction with impurities.
The variable $z$ in eqs.~(\ref{self-energy})--(\ref{eta}) can assume arbitrary
complex values. Description of spectroscopic experiments requires knowledge of
the retarded electronic self energy at real frequencies, which corresponds to
solutions with $z=\omega+{\rm i}0^{+}$. A fast and stable numerical procedure
for this purpose was proposed by Marsiglio {\it et al} \cite{marsiglio88}.
It starts with computing the solutions for $\Sigma (z)$ on the imaginary axis,
at $z={\rm i}\omega _{m}$, where eq.~(\ref{self-energy}) is simpler.
Then, the function $\eta ({\rm i}\omega _{m})$ is used to set up an
iterative procedure to find $\Sigma (\omega + i0^{+})$ just above the real axis.

The quantity $N_0(\xi)$ appearing in Eq.~(\ref{eta}) is the bare EDOS.
In this paper we use for it the following simple model:
\begin{equation}
N_{0}(\xi )=N_{0}\Theta \left( W/2-|\xi |\right),
\label{DOS-bare}
\end{equation}
where $W$ is the bare band width and $\Theta (x)$ is the step function.
The constant $N_{0}$ is fixed by normalization: $N_{0} = 1/W$.
In this paper we retain particle-hole symmetry for simplicity, with the
chemical potential at the center of the band, $\mu=0$.
In the clean case it was shown \cite{knigavko05} that the main
characteristic features  appearing in the electronic self energy and the
memory function due to the finite band width, do not depend significantly
on details of the bare electronic band.

The renormalized density of electronic states, or density of states
for quasiparticles, is defined by
\begin{equation}
N(\omega )=\int\nolimits_{-\infty }^{+\infty }d\xi N_{0}(\xi )A(\xi,\omega).
\label{dos-renorm}
\end{equation}
Here
\begin{equation}\label{spectral-weight}
A(\xi ,\omega )=-\mathop{\rm Im}G_{ret}(\xi ,\omega )/\pi
\end{equation}
is the electronic spectral density and the retarded Green's function
$G_{ret}(\xi ,\omega )$ is defined by the relation
\begin{equation}
\left[ G_{ret}(\xi ,\omega )\right] ^{-1}\equiv \left( \left[ G_{0}(\xi ,z)%
\right] ^{-1}-\Sigma (z)\right) _{z=\omega +i0^{+}} \label{Green-func-ret}
\end{equation}
with $G_{0}(\xi ,z)= 1/(z-\xi )$ being the free electron Green's
function.
The renormalized quasiparticle can be expressed in terms
of the function $\eta =\eta _{1}+{\rm i}\eta _{2}$ \ of Eq. (\ref{eta})
as follows $N(\omega )/N_{0}(0)=-\eta _{2}(\omega )/\pi$.

The renormalized density of states $N(\omega)$ is a very important
quantity. It features various signatures of the interaction of
electrons with phonons and impurities. Note that in the infinite
electronic band approximation with flat bare EDOS the renormalized
EDOS $N(\omega)$ remains constant and does not carry any physical
information. In the present context of a finite band, $N(\omega)$ for
electron--phonon system has been studied recently by Do\u{g}an and
Marsiglio \cite{dogan03} at $T=0$ and by Knigavko and Carbotte
\cite{knigavko05} at finite temperatures. Below we emphasize the
analysis of the combined effect of both phonons and impurities. We
find that knowledge of the features of the renormalized EDOS
helps to understand better the behavior of the other spectroscopic
quantities such as the memory functions, which are related to the
optical response.
The renormalized EDOS $N(\omega)$ itself is a measurable quantity
and can be directly
probed by tunneling spectroscopy or angle-integrated photoemission
spectroscopy \cite{reinert00,chainani00,reinert03}. The accuracy of
the latter technique has increased dramatically in recent years and
properties of both new and traditional materials have been
scrutinized. It has been argued in Ref.~\onlinecite{knigavko05-2}
that normal state boson structure should be detectable in such
experiments for metals with electronic band width of order a few eV.

Let us return to the electronic self energy.
For the purpose of the following discussion we present the general
Eq. (\ref{self-energy}) for the values of the argument just above
the real axis, namely $z=\omega +i 0^{+}$ and at temperature $T=0$
(note that henceforth we will use real axis variable, such as
$\omega$, as shorthand for $\omega +i0^{+}$). Separating the real
and imaginary parts of $\Sigma$ we obtain the following expressions:
\begin{eqnarray}
  \Sigma_1(\omega) &=& \Gamma \,
  P\!\!\int_{-\infty}^{\infty} \frac{d\omega'}{\omega-\omega'} \frac{N(\omega')}{N_0(0)}
  + 2\omega \, P\!\!\int_0^{\infty} d\omega' \frac{N(\omega')}{N_0(0)}
   \int_0^{\infty} d\Omega \frac{\alpha^2F(\Omega)}{\omega^2-(\omega'+\Omega)^2} ,
  \label{self-energy-RE} \\
  \Sigma_2(\omega) &=& -\pi \left[
  \Gamma \frac{N(\omega)}{N_0(0)} +\int_0^{\omega} d \Omega
  \alpha^2F(\Omega) \frac{N(\omega-\Omega)}{N_0(0)} \right],
  \label{self-energy-IM}
\end{eqnarray}
where the symbol $P$ in Eq.~(\ref{self-energy-RE}) means that in the divergent
integrals over $\omega'$ the Cauchy principal value has to be taken.
Note that the self consistent nature of this equations is now masked. On the
right hand side of Eqs.~(\ref{self-energy-RE}) and (\ref{self-energy-IM}) the
self energy enters only via the renormalized EDOS $N(\omega)$.

The quasiparticle mass renormalization is defined by the relation:
\begin{equation}
\lambda^{(eff)}_{qp} =
-\lim_{\omega\rightarrow 0} {\rm d}\Sigma_1{(\omega}) / {\rm d}\omega .
\label{lam-eff-qp-def-1}
\end{equation}
From Eq.~(\ref{self-energy-RE}) we find that it is given by
\begin{eqnarray}
\lambda^{(eff)}_{qp} &=&
\Gamma \, P\!\! \int_{-\infty}^{\infty}
\frac{d\omega}{\omega} \frac{N'(\omega)}{N_0(0)}
+2 \int_0^{\infty} d \Omega \alpha^2F(\Omega)
\int_0^{\infty} \frac{d\omega}{(\omega+\Omega)^2} \frac{N(\omega)}{N_0(0)},
  \label{lam-eff-qp-expr-1}
\end{eqnarray}
where $N'(\omega)\equiv {\rm d} N(\omega)/{\rm d}\omega$ is the derivative of
the renormalized EDOS.
For an infinite band with a constant $N(\omega)$ we recover the known result.
The second term on the right hand side of Eq.~(\ref{lam-eff-qp-expr-1})
reduces to $\lambda=2\int_0^\infty d\Omega \alpha^2F(\Omega)/\Omega$,
the usual expression for the mass renormalization due to electron--phonon
interaction, while the first term, which represents the effect of the elastic
scattering from impurities, vanishes. In the case of an energy dependent EDOS
impurities produce a finite contribution to the quasiparticle mass
renormalization. Our subsequent numerical analysis shows that for a finite band
$N'(\omega)$ is a decreasing function of $\omega$ for $\omega>0$ in large intervals,
which become especially substantial if $\omega_D \ll W$ [remember that
$N(-\omega)=N(\omega)$ because we consider the half filling case].
This makes the impurity contribution to
$\lambda^{(eff)}_{qp}$ negative and opposite in sign to the phonon contribution.
Therefore, in a finite electronic band the increased elastic scattering results
in the apparent decrease of the magnitude of the electron--phonon interaction,
as specified by $\lambda^{(eff)}_{qp}$.

In this paper we solve the self consistent equations for the self energy
numerically. To better understand the trends observed in  our
numerical results it is helpful to have analytic, though maybe not
exact, expressions for the self energy. We found that a useful approximation
is to replace the renormalized EDOS $N(\omega)$ in Eq.~(\ref{self-energy-RE}),
(\ref{self-energy-IM}) and (\ref{lam-eff-qp-expr-1}) with the bare EDOS
$N_0(\omega)$, given in the model we consider by a constant equal to $N_0(0)$
with cutoff at the bare band edge $W/2$ (see Eq.~(\ref{DOS-bare})).
This approximation amounts to disregarding the self consistency, while keeping
track of the finite width of the band.
It is expected, and we confirm this in the following Section, that this
approximation is good at small frequencies $\omega$ as long as the characteristic
phonon frequency is much smaller than the band width $W$.
Moreover, this non selfconsistent approximation allows us to obtain a simple estimate
for the characteristic frequency $\bar{\omega}_{qp}$ of a finite electronic band,
when the real part of the self energy $\Sigma_1(\omega)$ changes sign.
Deficiencies of the non selfconsistent approximation are discussed later.
The non selfconsistent results have the form:
\begin{eqnarray}
 \Sigma_1^{(ns)}(\omega)&=& \Gamma \ln\left|\frac{\omega+W/2}{\omega-W/2} \right|
+\int_{0}^{\infty }{\rm d}\Omega \,\alpha ^{2}F(\Omega ) \ln \left|
\frac{\omega-\Omega}{\omega+\Omega}\,
\frac{\omega+W/2+\Omega}{\omega-W/2-\Omega} \right|,
\label{sigma1-non-self} \\
 \Sigma_2^{(ns)}(\omega) &=& -\pi \left[ \Gamma \Theta(W/2-|\omega|)
  + \int_{0}^{\omega }{\rm d}\Omega \,\alpha ^{2}F(\Omega )
  \Theta(W/2-|\omega-\Omega|) \right],
  \label{sigma2-non-self} \\
  \lambda^{(ns)}_{qp} &=& - 2 \frac{\Gamma}{W/2} +2 \int_0^{\infty} d \Omega
\frac{\alpha^2F(\Omega)}{\Omega} \frac{1}{1+\Omega/(W/2)}.
  \label{lam-non-qp-expr}
\end{eqnarray}
These equations, while not exact, reduce to the well known infinite band
results in the limit $W\rightarrow\infty$ and show the modifications
brought about by a finite band. Note in particular that when
no impurities are present the mass renormalization is always reduced over
its infinite band value and that the quasiparticle scattering rate
$\tau^{-1}_{qp}(\omega)\equiv -2\Sigma_2(\omega+i 0^{+})$ drops to zero for
$\omega > W/2 + \omega_D$ instead of remaining constant as in the
infinite band case (we remind that $\omega_D$ denotes the maximum phonon
frequency in $\alpha ^{2}F(\Omega )$).

\subsection{Optical response}

For characterization of the optical response the quantity of interest is
the memory function, which is the optical counterpart of the self energy.
The memory function $M(\omega)=M_1(\omega)+iM_2(\omega)$ appears explicitly
in the following expression for the complex optical conductivity
$\sigma(\omega) =\sigma_{1}(\omega)+ {\rm i}\sigma_{2}(\omega)$:
\begin{equation}
\sigma (\omega )=
\frac{2S}{\pi }\frac{1}{M(\omega)-i\omega},
\label{extended-drude-band}
\end{equation}
which is also called in the literature the extended Drude formula
\cite{shulga1,allen1,optics-1,optics-2,tanner1}. In this equation
$S$ is the optical sum defined by the integral
\begin{equation}
S=\int_{0}^{+\infty }\sigma _{1}(\omega )d\omega .
\label{optical-sum}
\end{equation}
For the infinite and flat electronic band it was shown by Allen\cite{allen04}
that in the limit $\omega\rightarrow 0$ the memory function and the self
energy are closely related, namely:
$M_1(0)=-2\Sigma_2(0)$ and $M_2(0)=\Sigma_1(0)$. This is one of
the reasons why the real part of the memory function can be identified
as the optical scattering rate, $M_1(\omega)\equiv\tau _{op}^{-1}(\omega )$.
On the other hand, the imaginary part of the memory function is usually
related to the frequency dependent optical mass renormalization,
$M_2(\omega)\equiv -\omega \lambda_{op}(\omega )$. The memory function
can easily be found if the conductivity is known. Indeed from
Eq.~(\ref{extended-drude-band}) it follows that
\begin{eqnarray}
\tau _{op}^{-1}(\omega ) &=&\frac{2S}{\pi }\frac{\sigma _{1}(\omega )}{%
\sigma _{1}^{2}(\omega )+\sigma _{2}^{2}(\omega )},
\label{mem-func-RE} \\
-\omega \lambda _{op}(\omega ) &=&\omega -\frac{2S}{\pi }\frac{\sigma
_{2}(\omega )}{\sigma _{1}^{2}(\omega )+\sigma _{2}^{2}(\omega )}
\label{mem-func-IM}
\end{eqnarray}
and these are the relations that we used in our numerical work presented below.

The optical conductivity was obtained using linear response theory, neglecting
vertex corrections. The details were described previously
elsewhere \cite{knigavko05}, and here we just write down
the expression for the real part of the conductivity:
\begin{eqnarray}
\sigma _{1}(\omega )
&=&
\frac{2\pi e^{2}}{\hbar ^{2}}
\int_{-\infty }^{+\infty}{\rm d}\xi N_v(\xi )
\int_{-\infty }^{\infty }{\rm d}\omega^{\prime }
A\left( \xi,\omega^{\prime }\right)A\left( \xi ,\omega^{\prime } +\omega \right)
\frac{f\left( \omega^{\prime } \right) -f(\omega^{\prime } +\omega )}{\omega },
\label{cond-real-part}
\end{eqnarray}
and point out that the corresponding imaginary part was computed
as the Hilbert transform of the real part, based on the the Kramers--Kronig
relations. In Eq.~(\ref{cond-real-part})
$N_v(\xi)\equiv N_{0}(\xi )v_{\xi }^{2}$ where $v_{\xi }^{2}$ is the
averaged over the Brillouin zone the square of the group velocity
defined for a general dispersion in Ref.~\onlinecite{marsiglio90}.
In the following discussion of the optical response and in the
numerical calculations of this paper we assume that the
system is isotropic and use for $v_{\xi }^{2}$ the expression
$v_{\xi }^{2}=\frac{2\hbar ^{2}}{mD}\left( \frac{W}{2}+\xi \right)$,
derived from the quadratic dispersion of free electrons with lower
band edge at $\xi =-W/2$ ($D$ is the number of spatial
dimensions, $m$ the free electron mass).
It is useful to introduce the optical effective mass renormalization as
\begin{equation}
\lambda^{(eff)}_{op} = \lambda_{op}(\omega=0),
\label{lam-eff-op}
\end{equation}
which is the quantity to be compared with the quasiparticle effective mass
renormalization $\lambda^{(eff)}_{qp}$ of Eq.~(\ref{lam-eff-qp-def-1}).
We will see that complete numerical results indicate that these
two quantities are not the same in a finite band. Another important quantity
that we will discuss below is the optical scattering rate at the Fermi level,
$\tau _{op}^{-1}(\omega=0)$.

As we have done for the self energy it is helpful in understanding the
complete numerical results, that will be presented in the following
sections, to have simple although approximate analytic expressions
for the optical quantities with which to compare. It is not feasible
to obtain a simple accurate expression for the optical scattering rate
in the general case, and even our model of bare EDOS with sharp cutoffs
(see Eq.~(\ref{DOS-bare}))
do not provide enough simplifications for this purpose. We decided therefore
to make use of the existing expressions, valid for the infinite band.
Historically, based on second order perturbation theory
for the electron--phonon system Allen \cite{allen71} was the first to
provide such an equation valid at zero temperature in the infinite flat band case.
A generalization to finite temperature was made by Shulga {\it et al}
\cite{shulga91} using a very different method which starts with the
Kubo formula and makes approximations to get the same result as Allen
when $T\rightarrow0$ limit is taken.
On the other hand Mitrovi\'c and Fiorucci \cite{mitrovic85}
and Mitrovi\'c and Perkowitz \cite{mitrovic84} have generalized Allen's
original work to include the possibility of an energy dependent
electronic density of states. They considered only zero
temperature. Very recently Sharapov and Carbotte \cite{sharapov05}
have provided a finite temperature extension based on the Kubo formula.
Such formulas have also been used recently in analysis of data
\cite{hwang05,dordevic05} and in comparison with more complete
approaches \cite{carbotte05,schachinger03}. The formulas for
the optical effective mass and scattering rate, which are
suitable for our forthcoming discussion, are those for $T=0$. They are
given in Refs.~\onlinecite{mitrovic85} and \onlinecite{mitrovic84}
and here we reproduce them for the reader's convenience:
\begin{eqnarray}
  \lambda_{op}(\omega) &=& \frac{2 \Gamma}{\omega^2}
    P\int_0^\infty {\rm d}\omega' \frac{N(\omega')}{N_0(0)}
    \ln\left| \frac{\omega'^2}{\omega'^2 - \omega^2}\right|
  + \frac{2}{\omega^2}
  \int_{0}^{\infty }{\rm d}\Omega \,\alpha ^{2}F(\Omega )
  P\int_0^\infty {\rm d}\omega' \frac{N(\omega')}{N_0(0)} \ln\left[
  \frac{(\omega'+\Omega)^2}{(\omega'+\Omega)^2-\omega^2} \right],
  \label{lam-inf-band}\\
    \tau_{op}^{-1}(\omega) &=&  \frac{2 \pi \Gamma}{\omega}
    \int_0^\omega {\rm d}\omega' \frac{N(\omega')}{N_0(0)}
  + \frac{2\pi}{\omega}
  \int_{0}^{\infty }{\rm d}\Omega \,\alpha ^{2}F(\Omega )
  \int_0^{\omega-\Omega} {\rm d}\omega' \frac{N(\omega')}{N_0(0)},
  \label{tauINV-inf-band}
\end{eqnarray}
where the symbol $P$ means, as usual, that the $\omega'$ integrals
are calculated as the principal Cauchy values. In the above equations
$N(\omega)$ is the renormalized EDOS that fully incorporates
the self consistent electronic self energy. To grasp the finite band
effects in the memory function we intend to replace $N(\omega)$
with the bare EDOS $N_0(\omega)$ from Eq.~(\ref{DOS-bare}) similarly
to our approach to the derivation of
Eqs.~(\ref{sigma1-non-self})--(\ref{sigma2-non-self})
for the non self consistent self energy.
Note however that in order to arrive at Eqs.~(\ref{lam-inf-band})
and (\ref{tauINV-inf-band}) it is necessary to assume
\cite{mitrovic85} that $N_v(\xi)$ is constant and extends to
infinity, i. e. no cut off is applied to it. This means that after
the proposed replacement effectively only finite band effects originating
in the self energy are included in the optical quantities and the resulting
approximate formulae are not expected to be quantitatively correct
for all frequencies $\omega$. Nevertheless we found that such analytical
expressions are very useful at $\omega<W/2$. They read:
\begin{eqnarray}
  \lambda_{op}(\omega) &=& \frac{2\Gamma}{\omega}
  \left[  \ln \left|\frac{W/2-\omega}{W/2+\omega}\right|
  -\frac{W/2}{\omega}\ln\left|1-\left(\frac{\omega}  {W/2}\right)^2\right|
  \right]
  +\frac{2}{\omega}
  \int_{0}^{\infty }{\rm d}\Omega \,\alpha ^{2}F(\Omega )\left[
  \ln\left|\frac{\omega-\Omega-W/2}{\omega+\Omega+W/2}\,
  \frac{\omega+\Omega}{\omega-\Omega}\right| \right.
  \nonumber \\
  && - \left. \frac{\Omega}{\omega}
  \ln \left|\frac{\Omega^2}{\omega^2-\Omega^2}\,
  \frac{\omega^2-(\Omega+W/2)^2}{(\Omega+W/2)^2}\right|
  +\frac{W/2}{\omega}\ln\left|\frac{(\Omega+W/2)^2}
  {\omega^2-(\Omega+W/2)^2}\right| \right],
  \label{lam-fin-band-1}\\
  \tau_{op}^{-1}(\omega) &=&
  2\pi\Gamma\left|\Theta(W/2-\omega)+ \frac{W/2}{\omega}\Theta(\omega-W/2)\right|
  \nonumber \\
  && + \frac{2\pi}{\omega}
  \int_{0}^{\infty }{\rm d}\Omega \,\alpha ^{2}F(\Omega )
  \Theta(\omega-\Omega)\left[(\omega-\Omega)\Theta(W/2-(\omega-\Omega))
  +W/2\Theta(\omega-\Omega-W/2)\right].
  \label{tauINV-fin-band-1}
\end{eqnarray}
In particular, Eq.~(\ref{lam-fin-band-1}) is used below to obtain a
reasonable estimate for the characteristic frequency $\bar{\omega}_{op}$
at which the imaginary part of the memory function changes sign.

\section{Results due to a band cutoff}
Motivated by the electron--phonon interaction in the fulleride compound
K$_{3}$C$_{60}$, we use a three frequency model for the electron-phonon
spectral function:
\begin{equation}
\alpha ^{2}F(\omega )=\lambda \sum_{i=1}^{3}\frac{\omega _{i}l_{i}}{2}\delta
\left( \omega -\omega _{i}\right)
\label{three-freq-model}
\end{equation}
with $\sum_{i=1}^{3}l_{1}=1.$ The interaction strength $a$ is
defined as the area under the $\alpha ^{2}F(\omega )$ curve. The
mass enhancement parameter $\lambda $ is given by eq.~(\ref{lambda})
with $z=0$. We set $\lambda =0.71$ with
$l_{1}=0.3,l_{2}=0.2,l_{3}=0.5$ and $\omega _{1}:\omega
_{2}:\omega_{3}=0.04:0.09:0.19$ eV
\cite{choi98}.
This model has $a=43.8$ meV and $\omega _{\ln }=102.5$ meV, where
$\omega _{\ln }$ is the logarithmic frequency \cite{mars-book}, a
convenient parameter to quantify the phonon energy scale. For the
forthcoming discussion we set $W=2.5$ eV, which leads to a small
value for the adiabatic parameter $\omega _{\ln }/(W/2)=0.082$.
For the model of Eq.~(\ref{three-freq-model}, the non self consistent
approximation to the self energy given by
Eqs.~(\ref{sigma1-non-self})--(\ref{lam-non-qp-expr}) becomes
\begin{eqnarray}
  \Sigma^{(ns)}_1(\omega) &=&
  \Gamma \ln\left|\frac{\omega+W/2}{\omega-W/2} \right|
  +\lambda \sum_{i=1}^3 \frac{\omega _{i}l_{i}}{2}
  \ln\left| \frac{\omega-\omega_i}{\omega+\omega_i}\,
  \frac{\omega+W/2+\omega_i}{\omega-W/2-\omega_i}\right|,
  \label{sigma1-non-self-3freq} \\
  \Sigma^{(ns)}_2(\omega) &=& -\pi \left[ \Gamma \Theta(W/2-|\omega|)
  +\lambda \sum_{i=1}^3   \frac{\omega _{i}l_{i}}{2} \Theta(|\omega|-\omega_i)
  \Theta(W/2-|\omega|+\omega_i) \right],
  \label{sigma2-non-self-3freq} \\
  \lambda^{(ns)}_{qp} &=& - 2 \frac{\Gamma}{W/2} +
  \lambda \sum_{i=1}^3 \frac{l_i}{1+\omega_i/(W/2)}.
  \label{lam-non-qp-expr-3freq}
\end{eqnarray}
For the parameters chosen the non selfconsistent mass renormalization is
$0.63$ in the clean case to be compared with $\lambda=0.71$.
\begin{figure}[t]
\centering
\includegraphics*[width=.5\textwidth]{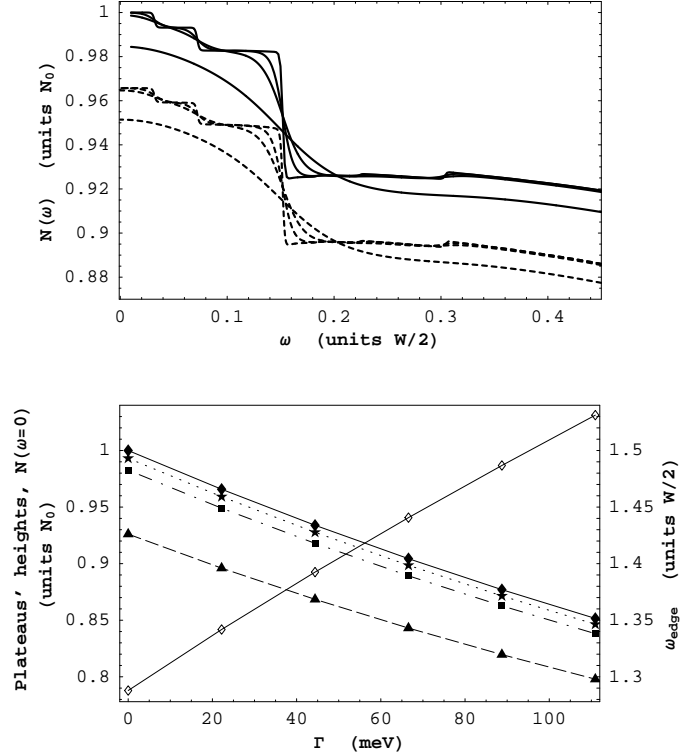}
\caption{ Top frame: dressed density of states $N(\omega)$ vs
$\omega$ for the $\alpha^2 F(\omega)$ model of
Eq.~(\ref{three-freq-model}) with $\lambda=0.71$ and the bare
bandwidth of $W=2.5$ eV. Impurity parameters are $\Gamma=0$ (solid)
and $22.2$~meV (dashed). In each group the temperatures are $14.5$,
$72.5$, $145$ and $435$ K (from top  to bottom).
Bottom frame: impurity dependence of various characteristic features
of renormalized density of states $N(\omega)$ at low temperature
$T=14.5$~K. Left vertical axis: dependence of phonon plateaus height
(starts, squares, triangles) on impurity parameter $\Gamma$ compared
with $N(\omega)=0$ vs $\Gamma$ dependence (diamonds). Right vertical
axis: $\omega_{edge}$ vs $\Gamma$.} \label{fig-dos-d5}
\end{figure}

We begin by reviewing effects of the electron--phonon interaction
due to a finite bandwidth which are seen in the EDOS. Some of the
features have been studied previously in
Refs.~\onlinecite{dogan03,knigavko05,knigavko05-2}. Here we want to
emphasize  impurity effects and give the comparison between non
selfconsistent and fully selfconsistent results. In
Fig.~\ref{fig-dos-d5} (top frame) we show the frequency dependence
of the renormalized quasiparticle density of states $N(\omega)$
based on the three frequency model of Eq.~(\ref{three-freq-model})
with $\lambda=0.71$ and a half band width $W/2=1.25$~eV. These
parameters are by no means extreme yet the deviations from the
infinite band case [$N(\omega)=1$ for all $\omega$] are substantial.
First, note that a three step phonon structure is clearly seen at
small $\omega$ in the lower temperature curves. The top set of four
curves (solid) are for $\Gamma=0$, no residual scattering, and the
four lower curves (dashed) are for $\Gamma=22.2$~meV or a residual
scattering rate of 140~meV for the chosen value of $W$. The
temperatures are $T=14.5, 72.5, 145$ and $435$~K. For the $435$~K
curve the thermal smearing is large but not for the others. As the
impurity scattering rate is increased the band width increases but
the phonon structures do not smear appreciably. Instead their
relative amplitude is slightly attenuated. Identifying the three low
frequency plateaus in $N(\omega)$ we plot, in the bottom frame of
Fig.~\ref{fig-dos-d5}, their heights (solid stars, squares, triangles,
refer to the left vertical axis) as a function of $\Gamma$ and compare
with the value of EDOS at $\omega=0$ (solid diamonds, refer to the
left vertical axis). All are reduced in magnitude with
increasing $\Gamma$ but the difference between the height of the
third plateau and $N(\omega=0)$ is changed much less. At the same
time the band broadens by about 25\% (open diamonds, refer to the
right vertical axis) with the
$\omega_{edge}$ given by the right hand scale in units of $W/2$ [see
heavy solid line Fig.~\ref{fig-dos-d1} for a plot of $N(\omega)$ vs
$\omega$ over a larger energy scale which shows the band edge]. The
plateaus just described do not exist in an infinite band. This also
holds true for the substantial temperature dependence of
$N(\omega=0)$ seen in the top frame of Fig.~\ref{fig-dos-d5} as well
as the thermal smearing of the phonon structure.

The main features of renormalized ($T=0$) quasiparticle density of states
just described can be understood qualitatively and even semi quantitatively
in the context of the non selfconsistent approach. Recall
(see Eq.~(\ref{dos-renorm})) that the renormalized EDOS is
\begin{equation}
\label{dos-renorm-2}
    N(\omega) = -\frac{N_0(0)}{\pi}\int_{-W/2}^{W/2} d\xi
    \frac{\Sigma_2(\omega)}{[\omega-\Sigma_1(\omega)-\xi]^2
    +\Sigma_2(\omega)^2}.
\end{equation}
We first note from Eq.~(\ref{sigma2-non-self}) that
in the clean case the imaginary part of the self energy is zero
for $\omega<\omega_1$, the first phonon energy in the model
for $\alpha^2F(\omega)$ of Eq.~(\ref{three-freq-model}).
Hence  the Lorentzian in Eq.~(\ref{dos-renorm-2}) becomes a delta
function and as a result $N(\omega)/N_0=1$. Once $\omega>\omega_1$
but still $\omega<\omega_2$ the imaginary part of the self energy
becomes $\pi a_1$ (with $a_i=\lambda l_i \omega_i /2$). If the
integral in Eq.~(\ref{dos-renorm-2}) were not cutoff at $W/2$ but
instead extended to infinity we would again get $N(\omega)/N_0=1$
and consequently the dressed density of states would remain unaffected
by the electron--phonon interaction. But the finite band cutoff
reduces the value of the integral in Eq.~(\ref{dos-renorm-2}) by
$2a_1/(W/2)$, i. e. by missing area under the Lorentzian beyond $W/2$.
(As $\omega$ increases the integrand is no longer symmetric between
positive and negative $\omega$ regions but we ignore this for
our rough estimate so that $2a_1/(W/2)$ is an upper limit.)
As $\omega$ increases the $a_i$ add until we come to the end of
$\alpha^2F(\omega)$, i.~e. $\omega=\omega_3$ in our three delta
function model. At this frequency our rough estimate for the reduction
in $N(\omega)$ is 10\% while the numerical calculations give 7.5\%
(see the bottom frame of Fig.~\ref{fig-dos-d5}). This difference is due
in part to the application of the self consistency and to our overestimate
 of the missing area under the Lorentzian of Eq.~(\ref{dos-renorm-2}).
 When impurities are added, a constant $\omega$ independent term is
 added to the imaginary part of the self energy. and so in the non
 self consistent approximation $N(\omega)$ would now be reduced by
 $2\Gamma/(W/2)$ at all frequencies. This expectation is in qualitative and
 even semi quantitative agreement with the results presented in
the bottom frame of Fig.~\ref{fig-dos-d5}. While self consistency
effects are on the whole small, they are responsible for the fact
that the lines in the bottom frame of Fig.~\ref{fig-dos-d5} are not
quite linear in $\Gamma$ and also not quite parallel to each other.
At much higher energies
beyond the phonon structure $N(\omega)$ drops to zero as most of the
Lorenzian in the integral of Eq.~(\ref{dos-renorm-2}) falls outside
of the range of integration. This occurs for two reasons. First,
the Lorenzian becomes centered outside the range of integration
and, second, its width becomes small. Recall that according to
Eq.~(\ref{sigma2-non-self}) $\Sigma_2(\omega+i 0^{+})$ is zero
for $\omega>W/2 +\omega_D$ in the non selfconsistent model
(pure limit).

Finally we return to the top frame of Fig.~\ref{fig-dos-d5} and
consider more closely temperature effects. To make our main point
it is sufficient to consider the $T=435$~K curves and the value
of $N(\omega)$ at $\omega=0$. At any finite temperature the
imaginary part of the self energy just above the real axis is given by
the expression \cite{mitrovic83,pickett82,mitrovic-thesis81}:
\begin{eqnarray}
    -\Sigma_2(\omega)&=& \pi\Gamma \frac{N(\omega)}{N_0(0)}
    +\pi \int_0^\infty d\Omega \alpha^2F(\Omega)
    \left[\frac{N(\omega-\Omega)}{N_0(0)}[n(\Omega)+f(\Omega-\omega)]
    +\frac{N(\omega+\Omega)}{N_0(0)}[n(\Omega)+f(\Omega+\omega)]\right],
    \label{self-energy-IM-finite-T}
    \end{eqnarray}
which follows from Eqs.~(\ref{self-energy})--(\ref{eta})
and its $\omega\rightarrow0$ limit is
\begin{equation}\label{w0-self-energy-IM-finite-T}
    2\pi\Gamma^{(eff)} \equiv -2\Sigma_2(\omega=0)=
    2\pi\Gamma \frac{N(0)}{N_0(0)}
    +2\pi \int_0^\infty d\Omega \frac{2\alpha^2F(\Omega)}{\sinh(\Omega/T)}
    \frac{N(\Omega)}{N_0(0)}.
\end{equation}
The presence of this scattering rate will lead to a drop in $N(\omega)$
of Eq.~(\ref{dos-renorm-2}) by approximately $2\Gamma^{(eff)}/(W/2)$,
or about 0.015 for these values of parameters, coming from the
inelastic scattering which is in semiquantitative agreement with the
numerical data. Also note the non linearity in these equations.
As $\Gamma$ increases, for example, $N(\omega=0)$ decreases and thus
the impurity contribution to $\Gamma^{(eff)}$
of Eq.~(\ref{w0-self-energy-IM-finite-T}) also decreases.
As temperature is increased the inelastic contribution to
$\Gamma^{(eff)}$ also increases and this
further reduces $N(\omega=0)$ and consequently the effect of the
impurity scattering. The two processes are no longer independent.
\begin{figure}[t]
\centering
\includegraphics*[width=.5\textwidth]{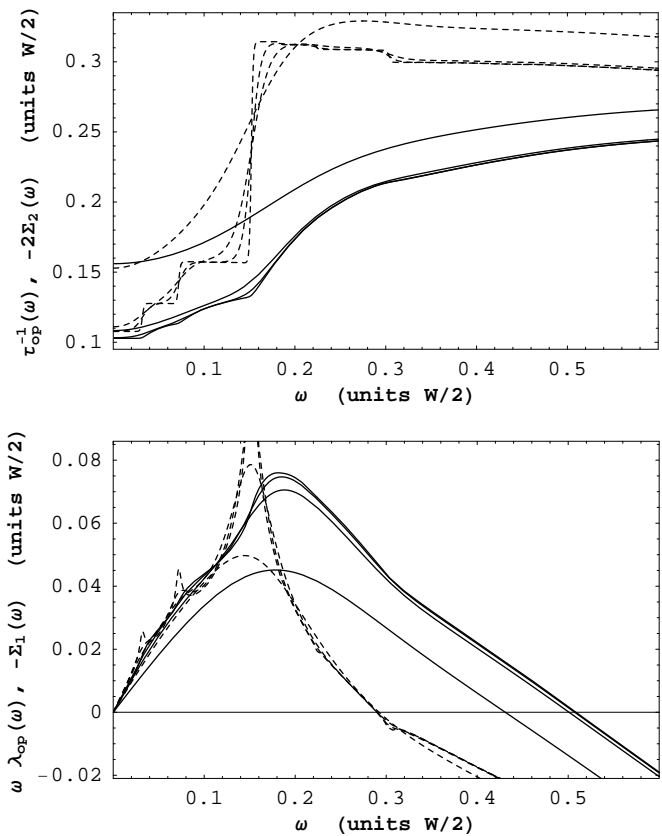}
\caption{ Top frame: the optical scattering rate
$\tau^{-1}_{op}(\omega)$  vs $\omega$ (solid) compared with the
quasiparticle scattering rate $\tau^{-1}_{qp}(\omega)$ vs $\omega$
(dashed) for $\Gamma=22.2$~meV. Temperature is
$T=14.5,72.5,145,425$~K from top to bottom at $\omega=0$. The bare
bandwidth is $W=2.5$ eV. Bottom frame: $\omega \lambda _{op}(\omega
)$ (solid) compared with negative of quasiparticle self energy
$-\Sigma _{1}(\omega )$ (dashed) for the same parameters as in the
top frame. } \label{fig-self-mem-d5}
\end{figure}

\begin{figure}[tp]
\centering
\includegraphics*[width=.5\textwidth]{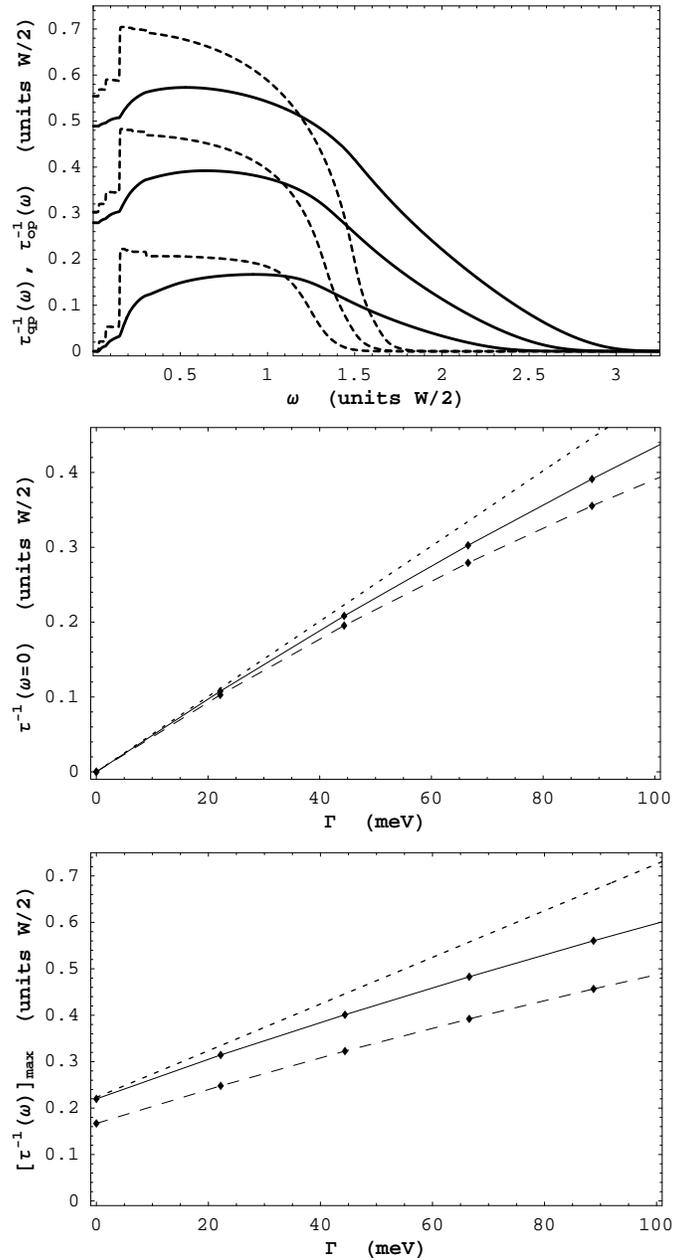}
\caption{Top frame: comparison of frequency dependence of optical
(solid) and quasiparticle (dashed) scattering rates at temperature
$T=14.5$~K. Impurity parameter $\Gamma=0, 67, 133$~meV (from bottom
to top at $\omega=0$). Middle frame: optical (dashed) and
quasiparticle (solid) scattering rates at $\omega=0$ vs $\Gamma$;
dotted line shows linear dependence $2\pi\Gamma$ for reference.
Bottom frame: Maximum value of optical (dashed) and quasiparticle
(solid) scattering rates vs $\Gamma$; dotted line shows linear
dependence $2\pi(a+\Gamma)$ for reference.} \label{fig-self2-tau-d5}
\end{figure}

\begin{figure}[tp]
\centering
\includegraphics*[width=.5\textwidth]{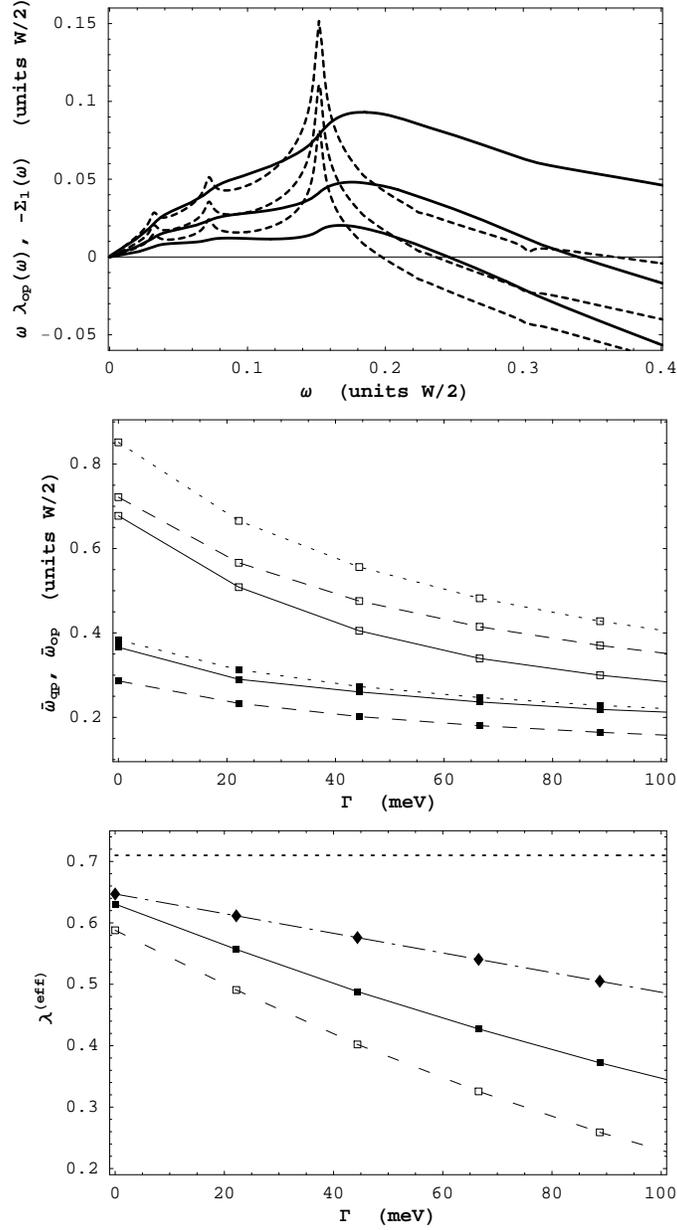}
\caption{ Top frame: comparison of frequency dependence of the
negative of the real part of quasiparticle self energy
$-\Sigma_1(\omega)$ (dashed curves) and corresponding memory
function $\omega\lambda_{op}(\omega)$ (solid curves) at temperature
$T=14.5$~K. Impurity parameter $\Gamma=0, 67, 133$~meV (from top to
bottom).
Middle panel: Frequency of zero crossing $\bar{\omega}$ vs $\Gamma$
for $\Sigma_1(\omega)$ (filled boxes) and $\lambda_{op}(\omega)$
(open boxes). Complete numerical results (solid curves) are compared
with non selfconsistent estimates based on the full
$\alpha^2F(\Omega)$ of Eq.~(\ref{three-freq-model}) (dotted curves)
and an appropriate Einstein oscillator spectrum (dashed curves).
Bottom frame: Complete numerical results for optical (dashed) and
quasiparticle (solid) effective mass renormalization of
Eqs.~(\ref{lam-eff-op}) and (\ref{lam-eff-qp-def-1}) vs $\Gamma$.
Dotted line refers to the input $\lambda=0.71$. Dash--dotted curve
shows the nonselfconsistent estimate of
Eq.~(\ref{lam-non-qp-expr-3freq}), equal for quasiparticle and
optical cases.} \label{fig-self1-omlam-d5}
\end{figure}

Of primary interest in this paper is the memory function of
Eqs.~(\ref{mem-func-RE})--(\ref{mem-func-IM}) also referred to as
optical self energy. In the top frame of Fig.~\ref{fig-self-mem-d5}
the optical scattering rate $\tau^{-1}_{op}(\omega)$ (solid) is
compared with the quasiparticle scattering $\tau^{-1}_{qp}(\omega)$
(dashed) given by $-2\Sigma_2(\omega)$, while in the bottom frame
$\omega \lambda_{op}(\omega)$ and $-\Sigma_1(\omega)$ are compared.
The parameters are the same as for Fig.~\ref{fig-dos-d5} but only
results for $\Gamma=22.2$~meV are presented. Many features of these
curves are worth notice. First, the three phonon steps in the
quasiparticle scattering rate (dashed), which are clearly seen in
the three lower temperature curves, are essentially wiped out for
$T=435$~K (uppermost curve). For this high temperature the inelastic
scattering due to collisions with thermally exited phonons has
substantially increased the value of $\tau^{-1}_{qp}$ at $\omega=0$
above the residual scattering. It has also lead to a qualitative
change in behavior at larger $\omega$. By contrast, the phonon
structures in the optical scattering rate at low temperature are
kinks rather than steps and therefore more difficult to identify.
Their temperature evolution is however very similar.
Complimentary to Fig.~\ref{fig-self-mem-d5}, in top frame of
Fig.~\ref{fig-self2-tau-d5} we compare the frequency dependence of
optical (solid curves) and quasiparticle (dashed curves) scattering
rates for several increasing values of the impurity parameter
$\Gamma$. Shown are the results for $\Gamma=0, 67, 133$~meV (from
the bottom to top) at the lowest temperature considered $T=14.5$~K.
In the top frame of Fig.~\ref{fig-self1-omlam-d5} we show similar
plots for the real part of the quasiparticle self energy (dashed)
compared with $\omega\lambda_{op}(\omega)$ (solid curves).

Note that, even for the lowest temperature considered in
Fig.~\ref{fig-self-mem-d5}, the residual scattering in both
quasiparticle and optical quantities are not exactly equal to their
infinite band values at zero temperature, which would be
$2\pi\Gamma$. In both cases it is smaller and also
$\tau^{-1}_{op}(\omega=0)<\tau^{-1}_{qp}(\omega=0)$. This difference
between a finite and an infinite band is further emphasized in the
middle frame of Fig. 3 where we have plotted $\tau^{-1}_{op}(\omega=0)$
(dashed curve), $\tau^{-1}_{qp}(\omega=0)$ (solid curve) as functions
of $\Gamma$ and compared with $2\pi\Gamma$ (dotted curve). The three
curves agree in the pure limit $\Gamma=0$ but the deviation between
these quantities increases as $\Gamma$ increases. The order remains,
with the optical scattering rate less than quasiparticle one, less
than the infinite band value $2\pi\Gamma$.
This behavior can easily be understood from the impurity contribution
to the imaginary part of the self energy of
Eq.~(\ref{self-energy-IM-finite-T}). As we have described before and
emphasize again, when $\Gamma$ increases $N(0)$ decreases so that
$2\pi\Gamma N(0)/N_0$ is less than $2\pi\Gamma$. The corresponding reduction
in $N(0)/N_0$ is roughly equal to $(1-2\Gamma/(W/2))$ which for $\Gamma=100$~meV
is about 16\% in good agreement with the solid curve of the middle frame
of Fig.~\ref{fig-self2-tau-d5} which gives the reduction in the
quasiparticle scattering rate  as the impurity $\Gamma$ is increased.
Note that the dashed curve for the corresponding optical quantity is even
lower than the quasiparticle one. This result comes from a full Kubo
formula calculation of the conductivity and is not captured by the
simplified formula of Eq.~(\ref{tauINV-fin-band-1}) for
$\tau^{-1}_{op}(\omega=0)$ which is $2\pi\Gamma N(0)/N_0$, the same as
for quasiparticles.  Note also that, as
temperature is increased and inelastic processes begin to contribute
to the scattering at $\omega=0$, $\tau^{-1}_{qp}(\omega=0)$ can become
smaller than $\tau^{-1}_{op}(\omega=0)$ but this order is reversed as
$\omega$ is increased (see $T=435$~K curve of
Fig.~\ref{fig-self-mem-d5}).

Finally we note that for higher temperatures the inelastic contribution
to $\Gamma^{(eff)}$ of Eq.~(\ref{w0-self-energy-IM-finite-T}) takes the
form:
\begin{equation}\label{gamma-high-T}
    \Gamma^{(eff)}(T)\sim
    2\int_0^\infty d\Omega \frac{\alpha^2F(\Omega)}{\Omega}
    \frac{N(\Omega)}{N_0(0)}\, T.
\end{equation}
This linear in temperature law is well known and the coefficient in
the square brackets would give the spectral lambda ($\lambda$) for
the infinite band case. For a finite band it is reduced as
$N(\Omega)/N_0(0))$ is less then one for all $\Omega$. We point out
that as the range of $\alpha^2F(\Omega)$, which is zero beyond $\omega_D$,
is well below (W/2) impurities and temperature will reduce the value of the
proportionality coefficient in Eq.~(\ref{gamma-high-T}) below its
$\Gamma=0,T=0$ effective value.

A second feature of the scattering rates shown in the upper panel of
Fig.~\ref{fig-self2-tau-d5}, which needs to be emphasized, are their
maximum values as a function of $\omega$. In the infinite band case
both the quasiparticle and optical scattering rates would rise to
the same asymptotic value at large $\omega$ which would be $2\pi a +
2\pi\Gamma$ at $T=0$. This expectation is modified by the
finite band cutoff.  In the bottom frame of
Fig.~\ref{fig-self2-tau-d5} we have plotted the maximum of
$[\tau^{-1}(\omega)]_{max}$ for optical (dashed) and quasiparticle
(solid) scattering rates as functions of $\Gamma$ and compared with
$2\pi a + 2\pi\Gamma$ (dotted). As can be seen from the top frame of
Fig.~\ref{fig-self2-tau-d5} the maximum in the quasiparticle rate
occurs at a frequency immediately above the third phonon step. For
the optical rate it occurs instead at much higher values of
$\omega$. For the pure case the frequency of the maximum in the
solid curve indeed falls beyond the bare band edge and is set not by the
value of the maximum phonon energy, but rather by the value of the
band edge itself. Also its maximum value is considerably smaller
than its quasiparticle counterpart [by about 25\%]. The deviation
between the two further increases with increasing impurity parameter
$\Gamma$. This difference between finite and infinite band results
has important implications for the analysis of experimental data.
Now the maximum in $\tau^{-1}_{op}(\omega)$ cannot be used as a
reliable estimate of the total area under the Eliashberg function
$\alpha^2F(\omega)$, often used as a measure of the electron--phonon
interaction strength. This is also the case for the quasiparticle
rate although the differences are not as substantial.
Note that the upper dashed curve in the top panel of
Fig.~\ref{fig-self2-tau-d5}, which gives the quasiparticle scattering
rate for $\Gamma=133$~meV, shows no flat region above $\omega_D$
as it has already started to drop due to band edge effects.
In fact, this is why it never reaches its infinite band maximum.
Such band edge effects are even more substantial for the
optical scattering rate which peaks only as $\omega\rightarrow\infty$
in the infinite band case.  In our non selfconsistent model of
Eq.~(\ref{tauINV-fin-band-1}) the maximum in $\tau^{-1}_{op}(\omega)$
will occur at $W/2+\omega_D$. At this point it will have a value of
approximately $2\pi a(1-\omega_{\ln}/(W/2))$. This represents
a roughly 15\% reduction over its infinite band value in
reasonable agreement with the numerical data of the lower frame of
Fig.~\ref{fig-self2-tau-d5}.

The main feature of the curves shown in Fig.~\ref{fig-self-mem-d5}
and \ref{fig-self2-tau-d5} (top frames)
that we have just described can be
understood approximately from the non selfconsistent formulas given
in the previous section. Starting with the top frame of
Fig.~(\ref{fig-self2-tau-d5}) the three sharp steps in
$\tau^{-1}_{qp}(\omega)$ and the extended nearly flat region beyond
are captured in Eq.~(\ref{sigma2-non-self}) as is the cutoff at
higher energies beyond $\omega=W/2+\omega_D$ with
$\omega_D$ being maximum phonon energy
(see Fig.~\ref{fig-self2-tau-d5}, top frame), while the
exact energy where $\tau^{-1}_{qp}(\omega)$ starts dropping
to zero is not captured since it is due to selfconsistency
that was not included. Similarly,
Eq.~(\ref{tauINV-fin-band-1}) allows us to understand the main
differences between quasiparticle (dashed curves) and optical
(solid curves) scattering rates. The optical scattering rate does
not jump abruptly to a value of $2\pi a_1$ at $\omega=\omega_1$
as $\tau^{-1}_{qp}(\omega)$ does but rather grows out of zero
gradually as $2\pi a_1 (1-\omega_1/\omega)$ for frequencies
in the range $\omega_1<\omega<\omega_2$. Additional contributions
enter at $\omega_2$ and $\omega_3$. On the other hand, for
$\omega>W/2+\omega_D$ the optical scattering rate
$\tau^{-1}_{op}(\omega)$ does not
fall off sharply but decreases towards zero as $1/\omega$
(see Fig.~\ref{fig-self2-tau-d5}, top frame). In the numerical
selfconsistent calculations it goes faster than this and then
vanishes exponentially\cite{knigavko05}, but this is not captured
by Eq.~(\ref{tauINV-fin-band-1}).

Next, we return to the bottom panel of Fig.~\ref{fig-self-mem-d5}
to discuss  the real part of the quasiparticle self
energy $\Sigma_1(\omega)$ [the negative of it is shown by dashed
curves] and its optical counterpart $\omega \lambda_{op}(\omega)$
(solid curves).
Perhaps the most striking feature of the real part of the self
energy as a function of frequency is that it changes sign with
increasing $\omega$, as noted in the work of Cappelluti and
Pietronero \cite{cappelluti03}. In this paper we find that the
corresponding memory function (or optical self energy) also has
a ``zero crossing'' in a finite band. We observe that the frequency
of the zero crossing $\bar{\omega}$ is larger in
the optics (about 0.5 for the parameters in this Figure) than for
the quasiparticle self energy (less than 0.3). While the
quasiparticle crossing is nearly independent of temperature in the
case shown, the optical one is not, dropping below 0.44 at $T=425$~K.

In the top frame of Fig.~\ref{fig-self1-omlam-d5} we show additional
results for three impurity content, namely $\Gamma=0, 67, 133$~meV.
The zero crossing in the memory function shifts progressively to
lower frequency with increasing $\Gamma$, and the magnitude of the
maximum value of $\omega\lambda_{op}(\omega)$ at low $\omega$
decreases correspondingly. For quasiparticles (dashed curves) the
trend is the same. For even higher values of $\Gamma$ than shown in
Fig.~\ref{fig-self1-omlam-d5} $\lambda_{op}(\omega)$ can become very
small at small $\omega$ and even be negative for all (positive)
frequencies. The results of our complete selfconsistent numerical
calculations on the zero crossing frequency $\bar{\omega}$ vs $\Gamma$
dependence are summarized in the middle panel of
Fig.~\ref{fig-self1-omlam-d5} by solid curves with either open symbols
(optics) or filled symbols (self energy).

The appearance of the phenomenon of ``zero crossing'' can be
qualitatively understood with the help of
the expression for the non selfconsistent self energy,
Eq.~(\ref{sigma1-non-self}), and memory function,
Eq.~(\ref{lam-fin-band-1}) at $T=0$.
A crude, but reasonable estimate can be obtained in both cases
using the Einstein model for the electron--phonon spectral function:
$\alpha^2F(\omega)=A\;\delta(\omega-\omega_E)$ with $\omega_E$ chosen
to be a characteristic frequency of the spectrum, for example $\omega_{ln}$.
For $\Sigma_1(\omega)$ and in the case $\omega_E,\bar{\omega}\ll W/2$
(but arbitrary $\Gamma$ and $A$) an explicit expression for the
solution for the frequency of zero crossing $\bar{\omega}$ can be found:
\begin{equation}
\label{zero-cross-qp-small-Om}
     \bar{\omega}_{qp}=\sqrt{\frac{\omega_E(W/2)}{1+\Gamma/A}},
\end{equation}
where the subscript $qp$ means "quasiparticle", i.~e.
pertinent to the self energy. It is shown in the middle panel of
Fig.~\ref{fig-self1-omlam-d5} by dashed curve with filled boxes.
For infinite band $\bar{\omega}_{qp}$
shifts to infinity and ceases to exist, as expected.
Note that $\bar{\omega}_{qp}$, given by this formula, become
smaller as impurity parameter $\Gamma$ increases.
This behavior explains the trend seen in the complete numerical
results (solid curve with filled boxes). Note that the Einstein
mode non selfconsistent estimate of the zero crossing frequency
produces an underestimate of the complete numerical result.

Similarly, for $\lambda_{op}(\omega)$
we use the non selfconsistent expression Eq.~(\ref{lam-fin-band-1})
with the Einstein mode $\alpha^2F(\omega)$ and in the limit
$\omega_E,\bar{\omega}_{op}\ll W/2$ obtain the following simple
equation for $\bar{\omega}_{op}$:
\begin{equation}
\label{zero-cross-op-small-Om}
    \frac{\bar{\omega}_{op}^2}{\bar{\omega}_{qp}^2}
    =2+\ln\frac{\bar{\omega}_{op}^2}{\omega_E^2},
\end{equation}
which shows that $\bar{\omega}_{op}\approx\sqrt{2}\bar{\omega}_{qp}$
with logarithmic accuracy. The resulting dependence of
$\bar{\omega}_{op}$ on $\Gamma$ is shown by dashed curve with open boxes.
Again, the general trend of the complete numerical result dependence on
$\Gamma$ (solid curve with open boxes) is obtained, but in this case of
the optical mass renormalization the Einstein mode non selfconsistent
estimate produces an overestimate of the exact $\bar{\omega}_{op}$.

The results of attempts to improve the non selfconsistent estimate for the
zero crossing frequency by including the full electron--phonon spectral function
$\alpha^2F(\Omega)$ of Eq.~(\ref{three-freq-model}) are given
by dotted curves. The improvement is significant in the case of the self energy
(dotted curve with solid boxes). This demonstrates that $\bar{\omega}_{qp}$
depends on the shape of the spectrum quite strongly and is not influenced much by
the self consistency. But this improvement came at a price: we do not have a
simple formula now. In the case of the optical mass renormalization,
inclusion of the full spectrum (dotted curve with open boxes) does not bring
improvements for $\bar{\omega}_{op}$, it even make the estimate worse as
compared to the Einstein mode spectrum. This reminds us again of the restricted
quantitative power of Eqs.~(\ref{lam-fin-band-1}) and
(\ref{tauINV-fin-band-1}), as was discussed previously in Section II B.
These formulae nevertheless provide a valuable qualitative guidance to the
complete numerical results, when used properly.

It is clear from Eqs.~(\ref{zero-cross-qp-small-Om}) and
(\ref{zero-cross-op-small-Om}) that the zero crossing frequency
contains information on the boson energy scale involved in the
scattering process.
Even though it is a slight digression from the present
discussion we would like to point out a possible application
of this finding. In their recent work on
optical conductivity in high $T_c$ cuprates Hwang, Timusk
and Gu \cite{hwang04} indeed have found a change in sign of the
optical self energy (not shown in their plots). The frequency
at which this occurs varies from compound to compound but
is of the order 6000~cm$^{-1}$ for their optimum and
overdopped samples and smaller, of order 4000--5000~cm$^{-1}$,
in underdopped samples. In recent work Markiewicz
\cite{markiewicz05} {\it et al} estimated that the typical
band width in the oxides is of the order 1.0 to 2.0~eV for
the dressed band with bare band structure results typically a
factor two larger. This would indicate from the present work
a boson exchange energy well above 150~meV. This is much larger
than $\omega_{\ln}$ for a phonon mechanism and is consistent
instead with spin fluctuations or marginal Fermi liquid model
\cite{schachinger98}.

Finally, turning to the position of the peaks in
$\omega\lambda_{op}(\omega)$ we note that in the infinite band case
it would fall at about $\sqrt{2}\omega_3$ for the model
$\alpha^2F(\omega)$ of Eq.~(\ref{three-freq-model}). Here finite
band effects have shifted it down by 15\% in the pure case. On the
other hand, including $\Gamma$ has little effect on peak's position
as can be seen in Fig.~\ref{fig-self1-omlam-d5}, top frame.

Another important characteristic of the curves in the top frame of
Fig.~\ref{fig-self1-omlam-d5} is the slope at $\omega=0$. For the
infinite band case it would give the mass enhancement parameter $\lambda$.
In the case of a finite electronic band this parameter can no longer
be directly read off the slopes because they are changed.
Denoting these by $\lambda^{(eff)}_{qp}$ and
$\lambda^{(eff)}_{op}$ in the cases of self energy and memory
function respectively (see Eq.~(\ref{lam-eff-qp-def-1}) for example),
we find that they are no longer equal to each other and are
sensitive to impurity content, as can be seen in the top frame of
Fig.~\ref{fig-self1-omlam-d5} where we plot $\omega\lambda_{op}(\omega)$
and $-\Sigma_1(\omega)$ vs $\omega$ for three impurity
content, namely $\Gamma=0, 67, 133$~meV, for a low temperature $T=14.5$~K.
They also depend on temperature as shown in the bottom frame of
Fig.~\ref{fig-self-mem-d5}, but this is not qualitatively different
from the infinite band case. The dependence of the two
$\lambda^{(eff)}$ on impurity parameter $\Gamma$ for this temperature
(with the other parameters the same as for Fig.~\ref{fig-dos-d5})
is detailed in the bottom frame of Fig.~\ref{fig-self1-omlam-d5}.
The dotted line shows the input $\lambda=0.71$ for reference.
Both optical (dashed) and quasiparticle (solid) mass renormalization
decrease substantially with increasing $\Gamma$ and the former quantity
is always smaller.
Note that in the non selfconsistent case Eq.~(\ref{lam-non-qp-expr})
applies to both $\lambda^{(eff)}_{qp}$ and $\lambda^{(eff)}_{op}$. This
dependence is also shown by long dash -- dotted curve for comparison.
While in the pure case it agrees well with the exact result for the
quasiparticles, it deviates substantially from the exact optical curve.
As $\Gamma$ increases the deviations increase although the trend is
properly given. This difference described goes beyond the approximations
that we used to obtain Eq.~(\ref{lam-non-qp-expr-3freq}) from the exact
expression of Eq.~(\ref{mem-func-IM}) based on the Kubo formula for the
optical conductivity. These approximations are clearly not very accurate
but because the resulting formulas are quite simple and analytic they are
nevertheless useful.

\begin{figure}[pt]
\centering
\includegraphics*[width=.5\textwidth]{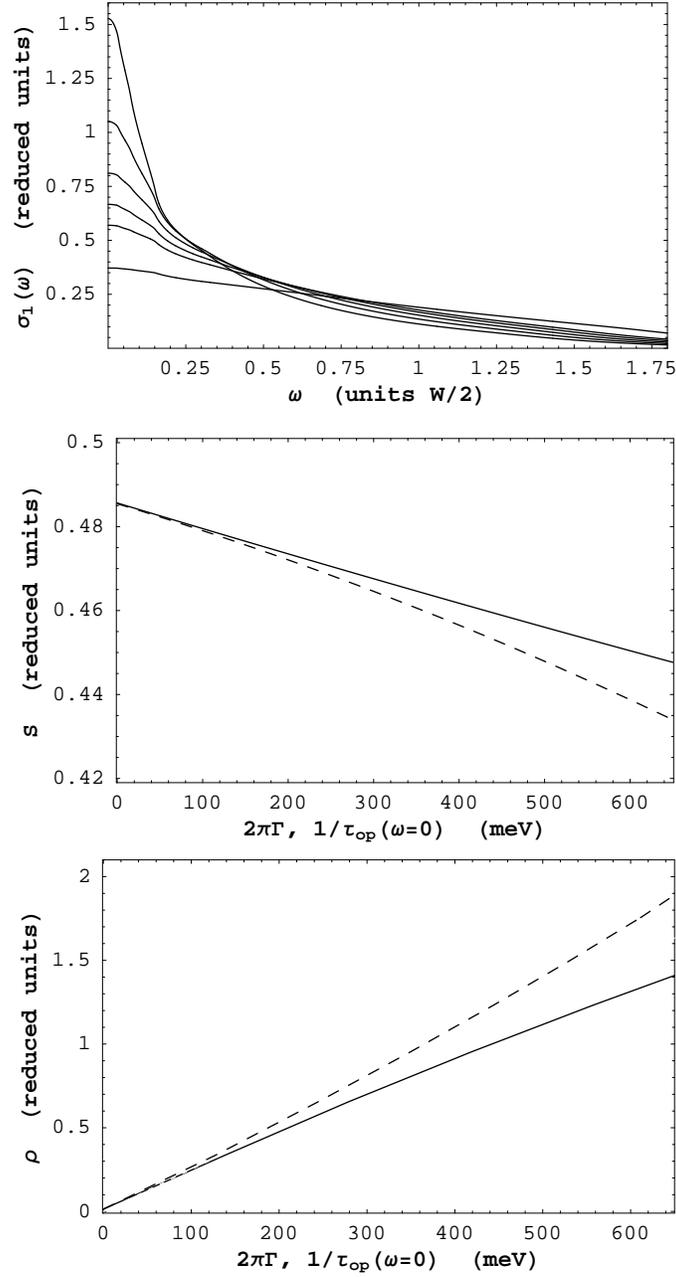}
\caption{ Top frame: real part of optical conductivity
$\sigma_1(\omega)$ vs $\omega$ for $\Gamma=44, 67, 89, 111, 133,
222$~meV (from top to bottom at $\omega=0$).
Middle frame: dependence of the optical sum $S$ of
Eq.~(\ref{optical-sum}) on $2\pi\Gamma$ (solid) and on
$\tau^{-1}_{op}(\omega=0)$ (dashed).
Bottom frame: dependencie of dc resistivity
$\rho=1/\sigma_1(\omega=0)$ on $2\pi\Gamma$ (solid) and on
$\tau^{-1}_{op}(\omega=0)$ (dashed). Dotted curve is the infinite
band result. Temperature is $T=14.5$~K for all frames.}
\label{fig-cond1-d5}
\end{figure}

In the top frame of Fig.~\ref{fig-cond1-d5} we show results for the
real part of the conductivity $\sigma_1(\omega)$ vs $\omega$
at temperature $T=14.5$~K.
Optical conductivity is measured in units $\pi e^{2}/[2Dm(W/2)]$.
Six values of impurity parameters are used, namely $\Gamma=44.4, 66.6,
88.8, 111.0, 133.1, 221.9$~meV. There are several features of these
curves which are different from corresponding infinite band
results. Perhaps the most obvious is that the total optical spectral
weight $S$, defined in Eq. (\ref{optical-sum}), is no longer
independent of temperature and impurity content.
[Temperature dependence is discussed in our previous
paper\cite{knigavko04}.]
The dependence of $S$ on $\Gamma$ is shown by the solid curve in
the middle frame of Fig.~\ref{fig-cond1-d5}.
The optical spectral weight is measured in units  $\pi e^{2}/[2Dm]$,
such that in the infinite band case $S=1/2$.
Because residual and inelastic scattering is no longer strictly
additive in finite bands, the impurity parameter $\Gamma$ cannot be
directly obtained from optical conductivity experiment.
Therefore, we also plot $S$ vs $\tau^{-1}_{op}(\omega=0)$
(dashed curve), a parameter that is measurable.
While there is a small difference between the
two plots, both show that the total optical spectral weight is
reduced as $\Gamma$ is increased. Another important property of the
conductivity $\sigma_1(\omega)$ from Fig.~\ref{fig-cond1-d5}, top
frame, which needs to be commented upon, is its dc value, or its
inverse $1/\sigma_1(\omega=0)$, the resistivity. This quantity is
plotted in the bottom frame of Fig.~\ref{fig-cond1-d5} where it is
seen to increase with $\Gamma$ (solid curve). This is also the case
when plotted with respect to $\tau^{-1}_{op}(\omega=0)$ (dashed curve),
and the dependence is not quite linear as would be expected in an
infinite band (dotted line).

The behavior of solid curve can be understood from the formula for
the resistivity:
$\rho(T)=[\pi/(2S)]\tau^{-1}_{op}(\omega=0)$.
While, as we have seen in a previous section, the approximate analytic
formulas that can be obtained for the optical quantities, are not as
accurate as for the self energy nevertheless they can be quite helpful
in providing insight into the complete numerical results. Sharapov
and Carbotte \cite{sharapov05} give the following expressions for
inelastic and impurity contributions, respectively:
\begin{eqnarray}
{\tau_{op,{\rm phon}}^{-1}(\omega\rightarrow0)}
&=& 4\pi \int_0^\infty d\Omega
\alpha^2F(\Omega) \int_{-\infty}^{+\infty} d\epsilon
\frac{N(\epsilon)}{N_0(0)}[n(\Omega)+f(\Omega-\epsilon)]
\left(-\frac{\partial f(\epsilon)}{\partial\epsilon}\right),
\label{tauINV-phon-inf}
\\
{\tau_{op,{\rm imp}}^{-1}(\omega\rightarrow0)} &=& 2\pi\Gamma
\int_{-\infty}^{+\infty} d\epsilon \frac{N(\epsilon)}{N_0(0)}
\left(-\frac{\partial f(\epsilon)}{\partial\epsilon}\right).
\label{tauINV-imp-inf}
\end{eqnarray}
Besides the explicit thermal factor appearing in these equations,
the temperature also enters through the renormalized DOS factor
$N(\epsilon)$ which also depends on the phonon $\alpha^2F(\Omega)$
and  on $\Gamma$ in contrast to the infinite band case.
Independent of the details, because the $N(\epsilon)$ factor is
everywhere smaller than its infinite band value of one, the
resistivity is always reduced below its infinite band value.
This reduction increase with increasing value of $\Gamma$ as we
see in lower frame of Fig.~\ref{fig-cond1-d5} and is also
increased with increasing temperature. At low temperatures the
appropriate measure of the decrease in the impurity term of
Eq.~(\ref{tauINV-imp-inf}) is the value of $N(0)$ while for
the inelastic term it depends on $N(\Omega)$ with $\Omega$
within the phonon range.

\section{Very narrow bands}

The case considered so far corresponds to a rather broad band as
compared with the phonon energy. Nevertheless we found important
qualitative changes from the infinite band case. Band
structure calculations\cite{satpathy92} for $K_3C_{60}$, as an example,
give a half band width of about $250$~meV  which is now
comparable to the energy of the maximum phonon energy of $190$~meV in
our model electron--phonon spectral density. For such cases
Kostur and Mitrovi\'{c} \cite{kostur93} and later
Pietronero, Str\"{a}ssler and Grimaldi \cite{pietronero95} have
considered the effect of vertex corrections and a generalization of
the Eliashberg equations which go beyond the Midgal theorem. The
specific case of the Pauli susceptibility was considered by
Cappelluti, Grimaldi and Pietronero \cite{cappelluti01}. In more
recent work, Cappelluti and Pietronero \cite{cappelluti03} recognized
that it was the effect of a finite band that primarily accounted for some
of the qualitative differences found in their previous work and proceeded
to include only these as a first step in understanding self energy
renormalization.
Here we follow their lead, but consider instead optical properties.

In this section we wish to accomplish three goals. First, we want to
understand differences that arise when very narrow bands are
involved as compared with relatively wide ones. Second, we want to
compare a case with a larger value of $\lambda$ and finally we replace
the three delta function model for $\alpha^2F(\omega)$ with a more
realistic extended spectrum. We consider a model with three
truncated Lorentzians:
\begin{equation}
\alpha ^{2}F(\omega )=
R(\lambda ) \sum_{i=1}^3 \frac{1}{2\pi }
\left[\frac{\delta_i }{(\omega -\omega_i )^{2}+\delta_1^{2}}
-\frac{\delta_i}{\eta_i^{2}+\delta_i^{2}}\right]
\Theta \left( \eta_i -\left| \omega
-\omega_i \right| \right) ,
\label{three-trunc-Lor-model}
\end{equation}
where $\omega_i$, the centers of the peaks, are the same as in
Eq.~(\ref{three-freq-model}). For each peak the parameter $\delta_i$
controls the half width, while the full spread is equal to
$2\eta_i.$ The rescaling factor $R(\lambda )$ is inserted to
guarantee a chosen value of $\lambda .$ Truncated Lorentzians are
often used in the literature to introduce a smearing of the simple
Einstein mode spectrum. For our numerical work
we picked $\delta_i=0.2\omega_i$ and $\eta_i=0.6\omega_i$. In this
case the peaks are wide and overlapping. The spectrum of
Eq.~(\ref{three-trunc-Lor-model}) has the characteristic
logarithmic \cite{mars-book} frequency of  $\omega_{ln}=96$~meV.

While, to set the parameters used here, we consider what might be
reasonable for $K_3C_{60}$, i. e. we chose $W/2=250$~meV and an
effective $\lambda$ of about one, we do not imply that our
calculations can be applied directly to this specific system. Other
complications such as the effect of Coulomb interactions \cite{yoo00}
may need to
be included as well. For example the small Drude peak seen in the
experiments of Degiorgi {\it et al} \cite{degiorgi94,degiorgi95}
which has a width of about $10-20$~meV and a weight representing
only about $10-20$\% of the total spectral weight is not understood
in our work. On the other hand qualitative features of the memory
function vs frequency dependence, such as the observed zero in its
real part, are captured.
\begin{figure}[t]
\centering
\includegraphics*[width=.65\textwidth]{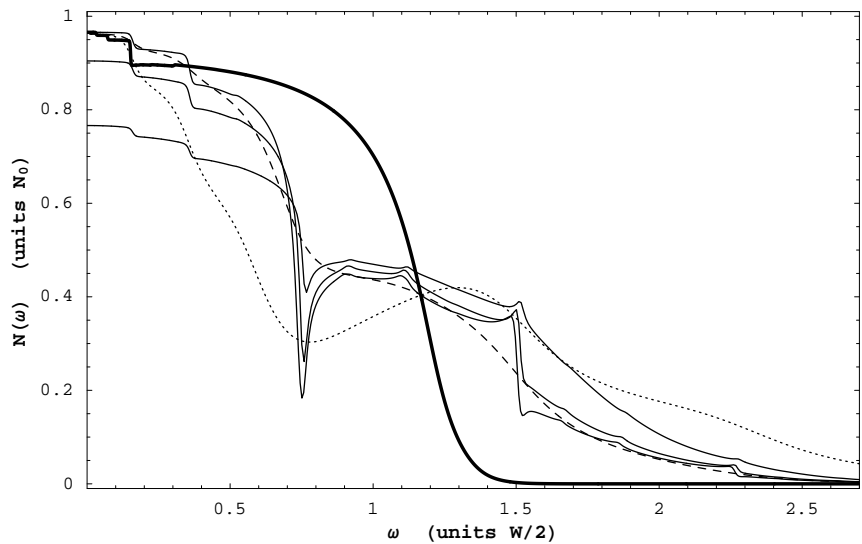}
\caption{ Frequency dependence of the renormalized density of states
$N(\omega)/N_0$ for a very narrow band with $W/2=250$~meV (thin
curves) with the parameters described in detail in the text. The
heavy curve refers to $N(\omega)/N_0$ with $W/2=1.25$~eV for
comparison.} \label{fig-dos-d1}
\end{figure}

In Fig.~\ref{fig-dos-d1} we present a series of results for the
dressed quasiparticle density of states $N(\omega)/N_0$ as a function of
energy $\omega$. The heavy solid curve shows previous results for a
rather wide band $W/2=1.25$~eV
on a broader scale for the three delta function model of
Eq.~(\ref{three-freq-model}) with $\lambda=0.71$ and
$\Gamma=22$~meV at low temperature $T=1.25$~K. It is to be compared
with the other curves all of which are for a band that is five times
narrower, namely $W/2=250$~meV. While
for the wider band significant phonon structures are limited to an
energy region well below the bare band cutoff at $\omega/(W/2)=1$, for the
narrower band they dominate the shape of $N(\omega)/N_0$ even beyond
$\omega/(W/2)=2$. No particular signature associated with the bare band
edge remains. This is distinct from the heavy continuous curve which
shows a smooth drop off at a new easily distinguishable renormalized
band edge energy increased somewhat over its bare value and smeared
by the interactions.

All thin solid curves in Fig.~\ref{fig-dos-d1} are for the three
delta function model of $\alpha^2F(\omega)$ with $\lambda=0.71$
at low temperature ($T=1.25$~K). The impurity parameter is
$\Gamma=4.4, 13.3$ and $39.9$~meV (from top to bottom). The
dashed curve correspond to the extended spectrum of
Eq.~(\ref{three-trunc-Lor-model}) with $\lambda=0.71$ and
$\gamma=5$~meV at low temperature ($T=2.15$~K).
The two sharp step like drops at $\omega/(W/2)<0.40$ present
in the delta function case are now almost
completely smeared out. The sharp spike like minimum at the energy
of the maximum phonon energy $\omega/(W/2)=0.76$ [$\omega=190$~meV]
and the near vertical drop at twice this energy, seen in the solid
curves are gone in the dashed curve as are the distinct multiphonon
structure at higher energies. The plateau like region at
$0.8<\omega/(W/2)<1.3$ [for $\omega$ between $190$ and $390$~meV]
in the solid curves becomes a shoulder in the dashed curve.

The modification of the renormalized density of states depends
on the mass renormalization parameter $\lambda$. To demonstrate this
we present in Fig.~\ref{fig-dos-d1} the result for the same extended
spectrum of Eq.~(\ref{three-trunc-Lor-model})) and the same impurity
parameter $\Gamma=4.4$~meV but with $\lambda=2$ (the dotted curve).
Now the low temperature step at $\omega/(W/2)=0.16$ becomes visible and a
deep and wide minimum develops in the frequency region of the most strongly
coupled part of the electron--phonon spectral density centered at
$\omega/(W/2)=0.76$. Additionally, the multiphonon processes become
stronger, which is manifested by appearance of the maximum at
$\omega/(W/2) \approx 1.4$ in the dotted curve in place of the shoulder in
the dashed curve. Finally, the renormalized band edge has been
shifted to much higher frequency and is not visible in
Fig.~\ref{fig-dos-d1}.
\begin{figure}[t]
\centering
\includegraphics*[width=.5\textwidth]{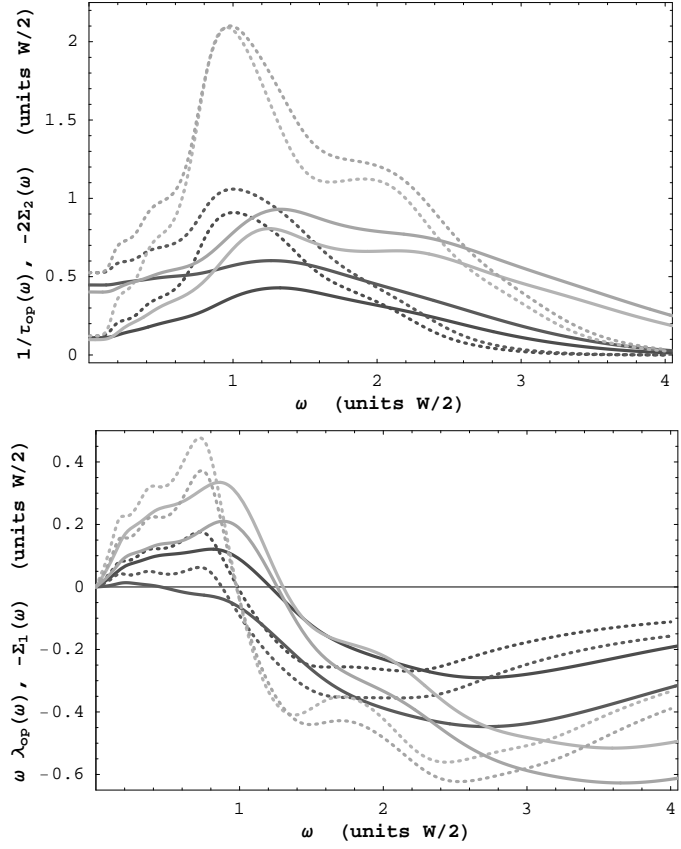}
\caption{ Top panel: Comparison of optical (solid curves) and
quasiparticle (dotted curves) scattering rates. Bottom panel:
Comparison of minus the real part of the self energy (dotted) with
the corresponding optical quantity $\omega\lambda_{op}(\omega)$
(solid). The mass renormalization is $\lambda=2.0$ (grey curves) and
$\lambda=0.71$ (black curves). In all cases there are two sets of
curves corresponding to impurity parameter $\Gamma=5$ and $25$~meV.
The extended $\alpha^2F(\Omega)$ of
Eq.~(\ref{three-trunc-Lor-model}) was used. The half band width
$W/2=250$~meV, temperature $T=2.15$~K.} \label{fig-self-mem-d1}
\end{figure}

Returning to the light continuous curves we note that, compared
to the wide band case of the previous section, the phonon steps are
now significantly reduced in magnitude as the impurity scattering
is increased. For example, compare the bottom curve for
$\Gamma=39.9$~meV to the top one for $\Gamma=4.4$~meV. These reductions
go beyond non selfconsistent approximation and demonstrate
that the selfconsistency becomes more important as the bare band
width is reduced. The near additivity of electron--phonon and
impurity contributions is lost. While for the curve corresponding
to the purest case, the non self consistent approximation predicts
well the size of the first step, it is not as good for the second
and the third step is quite off. This is expected as in this case
we are already not so far from the bare band edge and the
selfconsistency becomes essential.

Note that at $\omega=0$ for the bottom light solid line with
$\Gamma=39.9$~meV, $N(0)/N_0\approx 0.77$ in the numerical works. The
non selfconsistent estimate $(1-2\Gamma/(W/2))\approx 0.68$ which is
considerable smaller. However self consistency has actually
reduced the impurity scattering rate below its infinite band value
of $2\pi\Gamma$ because it is equal to  $2\pi\Gamma N(0)$ and
$N(0)/N_0(0)$ is smaller than one. Accounting for its 23\%
reduction eliminates much of the difference described above.
A similar semiquantitative argument can be made to understand
the reduction in phonon step size as $\Gamma$ increases when
the simple non selfconsistent estimate begin to fail. Finally,
we note the crossing of the light solid curves around
$\omega/(W/2)\approx 0.8$
and the increase in density of states in the tails beyond
$\omega/(W/2)=1.5$ as the impurity scattering is increased.

In Fig.~\ref{fig-self-mem-d1} we turn to the memory function
(solid lines) which is compared with the quasiparticle self energy
(dotted lines) for several cases.
We present results for $\lambda=0.71$ (black curves)
and for $\lambda=2.0$ (grey curves) and for two different impurity
content, $\Gamma=5$ and $25$~meV calculated with the model of
the extended $\alpha^2F(\Omega)$ of Eq.~(\ref{three-trunc-Lor-model}).
As expected both
quasiparticle and optical scattering rates rise to a higher maximum
value when $\lambda$ is larger, but the increase is not linear. This
is followed by a drop towards zero as $\omega$ gets large instead of
saturating at a common value of $2\pi a + 2\pi \Gamma$ as discussed
previously. The intercept of $\tau^{-1}_{op}(\omega)$ and
$\tau^{-1}_{qp}(\omega)$ at $\omega=0$ is related to the elastic
impurity scattering as the temperature for the figure is small
$T=2.15$~K. For $\tau^{-1}_{qp}$ it does not depend on the value of
$\lambda$ but for $\tau^{-1}_{op}$ it does. This can be seen most
clearly in the top set of curves for which $\Gamma=25$~meV but is
noticeable in the lower set with $\Gamma=5$~meV. When $\lambda$ is
larger $\tau^{-1}_{op}$ at $\omega=0$ is smaller.
This effect results from the Kubo formula and is not captured in any
of our approximate analytic formulas. As previously noted
$\tau^{-1}_{op}(\omega=0) < \tau^{-1}_{qp}(\omega=0) < 2\pi\Gamma$.

For $\lambda=2$ the maximum value of $\tau^{-1}_{qp}(\omega)$ (see
grey dotted curves in the top panel of Fig.~\ref{fig-self-mem-d1})
occurs at a frequency slightly below $\omega/(W/2)=1$ and is almost
the same for the $\Gamma=5$~meV and $\Gamma=25$~meV.
Although we have increased the impurity
scattering we have not gained in maximum quasiparticle scattering.
In an infinite band it would have risen from 3.22 to 3.72 for the
parameters used. This effect is due to the self consistency.
Both elastic and inelastic scattering involve not just
$\alpha^2F(\Omega)$ and $\Gamma$ respectively but also the self
consistent value of the dressed quasiparticle density of states
$N(\omega)/N_0$ which becomes reduced as $\Gamma$ is increased. This
in turn reduces both elastic and inelastic scattering rates leading,
in case considered, to a saturation of the maximum in
$\tau^{-1}_{qp}(\omega)$. For $\Gamma=5$~meV the rate is about 34\%
lower than its infinite band value and for $\Gamma=25$~meV it is
about 43\% lower. A similar situation holds for the case when
$\lambda=0.71$ (black dotted curves)
although in that case the maximum quasiparticle scattering does
increase slightly with increasing $\Gamma$, for $\Gamma=5$~meV it
is 26\% below its infinite band value and for $\Gamma=25$~meV it is
40\% below. Turning next to the optical scattering rate (solid
curves in Fig.~\ref{fig-self-mem-d1}, top frame), we note
first that they peaks at a higher frequency than does the
corresponding quasiparticle rates. Also they are considerably
smaller in magnitude, approximately 0.95 and 0.8 for $\Gamma=25$
and $5$~meV respectively for the case $\lambda=2$.

In the bottom frame of Fig.~\ref{fig-self-mem-d1} we compare
our results for the minus real part of the electronic self energy
$-\Sigma_1(\omega)$ (dotted curves) with the corresponding
optical quantity $\omega\lambda_{op}(\omega)$ (solid curves)
of Eq.~(\ref{mem-func-IM}),
which is the minus of the imaginary part of the memory function
$M(\omega)$ defined in Eq.~(\ref{extended-drude-band}).
It is quite clear that the optical masses (slopes at $\omega=0$
for solid curves) are in all cases considerably smaller than the
quasiparticle masses (slopes at $\omega=0$ for dotted curves).
As an example, for $\lambda=2$ the quasiparticle mass at
$\Gamma=5$~meV is 1.1 as compared to 0.70 for the optical mass,
while at $\Gamma=25$~meV we have 0.75 and 0.35 respectively.
Using the non self consistent  formula Eq.~(\ref{lam-non-qp-expr-3freq})
gives for both masses $\lambda=2$ ($\lambda=0.71$) 1.1 (0.39)
and 0.94 (0.21) for $\Gamma=5$~meV and $\Gamma=25$~meV respectively.
For the pure case the agreement for the quasiparticle mass is good
but this is no longer the case for $\Gamma=25$~meV. Also for the
purer case considered on Fig.~\ref{fig-self-mem-d1} the zero crossing of
quasiparticle and optical mass renormalization function can be
understood qualitatively with Eqs.~(\ref{sigma1-non-self})
and (\ref{lam-fin-band-1})  but these simple
estimates begin to fail for higher values of $\Gamma$.

Finally, we comment on the reflectivity data of Degiorgi {\it et al}
\cite{degiorgi94,degiorgi95} on $K_3C_{60}$.
They did not analyze their data to extract optical scattering rate
$\tau^{-1}_{op}(\omega)$ amd mass renormalization $\omega \lambda_{op}(\omega)$.
Nevertheless we infer from the data presented three qualitative
features. The value of $\tau^{-1}_{op}(\omega)$ at $\omega=0$ which gives
a measure of the residual scattering is of order 160~meV. This large value
is incompatible with the observed small Drude peak in $\sigma_1(\omega)$
 of width 20~meV which contains about 12~\% of the total optical
 spectral weight. Second, at $\omega\simeq500$~meV the scattering rate
 has increased to approximately 500~meV which implies an inelastic
 contribution of 360~meV. Such a rise is much larger than can be
 achieved in the model of Fig.~6 and would indicate that the bare band
width is somewhat larger than present band  structure calculations predict
and that the spectral $\lambda$ defined by the input electron--phonon
spectral function $\alpha^2 F(\omega)$ is even larger than 2.
Thirdly, $\omega \lambda_{op}(\omega)$ changes sign at approximately
220~meV after which it plunges towards large negative values. This feature
can be taken as the hallmark of finite band effects as it does not
occur in infinite band theories. While such a zero crossing occurs
naturally in our calculations and is generic, it is not clear that a
set of microscopic parameters chosen to reproduce the features of the
scattering rate $\tau^{-1}_{op}(\omega)$ would also accurately produce
the position of the zero in the optical mass. We did not attempt
such a combined fit as it would require a value of $\lambda$
which appears to be rather large.

\section{Conclusions}

In an electronic system with a constant bare electronic density of states $N_0$
the application of a finite band cut off profoundly modifies the
electron--phonon renormalization effects. In the infinite band case the dressed
quasiparticle density of states $N(\omega)$ remains equal to $N_0$ and is
independent of impurity scattering and temperature. A self consistent solution for
the self energy in a finite band shows that $N(\omega)$ acquires low energy
structure on the scale of the phonon energies. The band edge becomes smeared and
the band extends beyond the original bare cut off. This extension of the band to
higher energies is accompanied by a compensating reduction of spectral weight
below the bare cut off. Equally importantly $N(\omega)$ is affected by impurity
scattering and by temperature. $N(\omega=0)$ is reduced in both cases. On the other
hand, while temperature rapidly smears out the phonon structure, impurities mainly
reduce its amplitude.

The emphasis of previous works was on the effects of temperature
and impurity scattering on the electron self energy
$\Sigma(\omega)$, which is the quantity that determines
quasiparticle properties. Here we have extended these works to
optical properties and considered characteristic features of the
memory function. For the infinite band case elastic impurity
scattering just adds a constant amount ($2\pi\Gamma$) to the
quasiparticle inelastic scattering rate, but in our case, because of
the application of self consistency, they no longer add. Even at
$\omega=0$ we find that $\tau^{-1}_{op}<\tau^{-1}_{qp}<2\pi\Gamma$. The
well known result that, at high energies both optical and
quasiparticle scattering rates due to phonons become equal and
saturate at a value $2\pi a$ (with $a$ the area under the
electron--phonon spectral density) no longer holds.  While the
maximum in  $\tau^{-1}_{qp}$ can come close in value to $2\pi a + 2\pi
\Gamma$, the corresponding optical quantity $\tau^{-1}_{op}$ is much
smaller in magnitude. Its maximum value increases with increasing
$a$ but this increase is sublinear. A similar situation holds when
$\Gamma$ is increased. At yet higher energies both scattering rates
go to zero because of the finite band.

The known result that the real part of the quasiparticle self energy
is unaffected by the impurity scattering no longer holds and this
has an impact as well on the optical mass renormalization.
Quasiparticle and optical effective mass renormalization at $\omega=0$
now differ from each other, neither is equal to $\lambda$ and both depend
on impurity scattering. The real part of the self energy changes
sign with increasing energy as does  $\omega \lambda_{op}(\omega)$.
The energy at which the zero crossing occurs is set by the phonon energy
scale and can depend both on temperature and impurity content.
It is larger for optics than it is for the self energy in many cases
but not always. The optical spectral weight, i.~e. the area under the
absorptive part of the optical conductivity varies with temperature
and with impurity scattering. The elastic and inelastic contributions
to dc resistivity are no longer additive.

All these effects were found to be significant in magnitude even when
a rather modest value of mass enhancement parameter $\lambda=0.71$ is
used with a band width of 2.5~eV. While  a three delta function
$\alpha^2F(\omega)$ was used to emphasize boson structure with maximum
phonon energy of 190~meV, a broader spectrum was also considered. This
soften phonon structures but did not eliminate them. Of course, for
simple metals such as Pb, the band width is much wider than considered
above and the maximum phonon energy is also an order of magnitude
smaller, so that in this case the infinite band approximation is
appropriate and the finite band corrections found in this paper
would be negligible. We have found that increasing the value of $\lambda$
to 2 increases boson structure but the increase is not linear in the
value of $\lambda$. Also decreasing the value of W to 500~meV, a value
suggested by band structure calculations in the alkali doped C$_{60}$,
leads to a new regime in which important modifications due to the
electron--phonon interaction dominate at all energies and no easily
identifiable trace of the underlying bare electronic band cutoff remains.

While all these conclusions are based on numerical solution of the
selfconsistent equations for the self energy and the Kubo formula for
the conductivity, we have also derived more transparent analytical
formulas evaluated in a non selfconsistent approximation. These
simple formulas are not always accurate but give considerable
insight into complete numerical results obtained and prove valuable in
the analysis of optical data in finite band metals.

\section{Acknowlegements}

It is a pleasure to acknowledge useful discussions with J.~Hwang,
F.~Marsiglio, B.~Mitrovi\'{c} and S.G.~Sharapov. Work supported by
the Natural Science and Engineering Research Council of Canada
(NSERC) and the Canadian Institute for Advanced Research (CIAR).


\end{document}